\documentclass[fleqn,usenatbib,a4paper]{mnras}

\usepackage{graphicx}	
\usepackage{amsmath}	
\usepackage{amssymb}	

\usepackage{subcaption}
\title[Three-dimensional simulations of rapidly rotating core-collapse
 supernovae]{Three-dimensional simulations of rapidly rotating core-collapse
 supernovae: finding a neutrino-powered explosion aided by non-axisymmetric flows}

\author[T. Takiwaki et al.]{
Tomoya Takiwaki,$^{1,2}$\thanks{E-mail: takiwaki.tomoya@nao.ac.jp}
Kei Kotake$^{3,1}$ and Yudai Suwa$^{4,5}$\\
$^1$ Division of Theoretical Astronomy, National Astronomical Observatory of 
 Japan, 2-21-1, Osawa, Mitaka, Tokyo, 181-8588, Japan\\
$^2$ RIKEN, 2-1, Hirosawa, Wako, Saitama, 351-0198, Japan\\
$^3$ Department of Applied Physics, Fukuoka University, Jonan, Nanakuma, Fukuoka 814-0180, Japan\\
$^4$ Yukawa Institute for Theoretical Physics, Kyoto University,
 Oiwake-cho, Kitashirakawa, Sakyo-ku, Kyoto 606-8502, Japan\\
$^5$ Max-Planck-Institut f\"{u}r Astrophysik, Karl-Schwarzschild-Strasse 1,
D-85748 Garching, Germany
 }

\date{Accepted XXX. Received YYY; in original form ZZZ}

\pubyear{2016}

\begin{document}
\label{firstpage}
\pagerange{\pageref{firstpage}--\pageref{lastpage}}
\maketitle

 \begin{abstract}
We report results from a series of three-dimensional (3D) rotational
 core-collapse simulations for $11.2$ and 27 $M_{\odot}$ stars
 employing neutrino transport scheme by the 
isotropic diffusion source approximation.
  By changing the initial strength of rotation systematically,
  we find a rotation-assisted explosion for the 27$M_{\odot}$ progenitor
 , which fails in the absence of rotation. 
The unique feature was not captured in previous 
two-dimensional (2D) self-consistent rotating models because 
the growing non-axisymmetric instabilities 
play a key role. In the rapidly
 rotating case, strong spiral flows generated
 by the so-called low $T/|W|$ instability enhance
the energy transport from the proto-neutron star (PNS) to the gain 
region, which makes the shock expansion more energetic.
 The explosion occurs more strongly in the direction perpendicular to the 
 rotational axis, which is different from previous 2D predictions.
 \end{abstract}




\section{Introduction}\label{sec:introduction}
After a half-century of extensive research,
 theory and simulations are now converging to 
a point that multi-dimensional (multi-D) hydrodynamic instabilities 
 play a crucial role in the neutrino mechanism of core-collapse 
 supernovae (CCSNe, see \citet{thierry15,tony15,Janka12,burrows13,Kotake12_ptep} for reviews).
Multi-D fluid motions associated with convective overturn
 and the standing-accretion-shock-instability (SASI, \citet{blondin03})
 are the keys because buoyant and 
 turbulent flows increase 
the neutrino heating efficiency in the gain region, triggering the runaway 
expansion of the shock.
In fact, a growing number of neutrino-driven models
 have been recently reported in self-consistent two-dimensional (2D) 
simulations
, which has strengthened our confidence 
in the multi-D neutrino-driven paradigm 
(e.g., \citet{Marek09,suwa10,BMuller12b,dolence15,pan15,naka15,O'Connor15}).

This success, however, is now raising new questions.
With the exception in \citet{Bruenn13},
the explosion energies so far reported in these 2D models 
 are generally not sufficient to explain the canonical supernova 
kinetic energy of $10^{51}$ erg.
 Moreover, the most challenging self-consistent three-dimensional
 (3D) simulations with spectral neutrino transport have 
 failed to produce explosions for $11.2$, $20.0$, and 
$27.0 M_{\odot}$ progenitors \citep{Hanke13, tamb14}, or, in a few
 successful cases, showed much delayed neutrino-driven
 shock revival in 3D than in 2D (e.g., \citet{lentz15} and \citet{melson15b}),
 leading to even smaller explosion energies in 3D 
 compared to 2D (\citet{Takiwaki14}).

 One of the prime candidates to predominantly affect the 
CCSN explodability is general relativity (GR, e.g., \citet{BMuller12b,KurodaT12}).
Rotation (e.g., \citet{Marek09,suwa10}), 
magnetic fields \citep{endeve12,martin14}, 
and inhomogeneities in the progenitor core \citep{Couch15} 
are also attracting much attention to turn an unsuccessful multi-D model
 into a successful explosion.
 
In this {\it Letter}, we focus on
 the roles of rotation and 
 report results from a series of 3D rotational core-collapse simulations 
with spectral neutrino transport for $11.2$ and  $27.0 M_{\odot}$ stars. 
We briefly describe our numerical approach in Section \ref{model}. We
discuss our results in Section \ref{sec:results}, followed by a 
 summary in Section \ref{sec:summary} 

\section{Numerical Setup and Progenitor model}\label{model}

Initial conditions are taken from the $11.2$ and $27.0$ $M_{\odot}$ pre-supernova
 progenitors of \citet{Woosley02}.
 The models, which have been 
used in \citet{Takiwaki12,Takiwaki14,Hanke13,bernhard15},
 are useful to clearly explore the impacts of rotation
  The initially constant angular frequency of $\Omega_0 =1$ or $2$ rad/s 
 is imposed inside the iron core with a cut-off ($\propto r^{-2}$) outside.
Although these angular frequencies are close to the high-end of 
 those from most recent stellar evolution models 
(e.g., \citet{Heger00a,Heger05}, see also discussions in \citet{ott06}), 
 we assume such rapid rotation to clearly see the impacts of rotation 
in this study.
The model name is labeled as "s11.2-R1.0-3D", which represents
 the 11.2 $M_{\odot}$ model with $\Omega_0 = 1$ rad/s that 
 is computed in 3D simulation.  

 Our numerical method is based on that in \citet{Takiwaki14} except several points. 
We use the equation of state (EOS) by \citet{LSEOS}
(incompressibility $K=220$ MeV).  
Our code employs a 
high-resolution shock capturing scheme with an 
approximate Riemann solver of \citet{hlle} (see \citet{naka15} for more details).
For the calculation presented here, self-gravity is computed by a Newtonian monopole 
approximation\footnote{Our 3D rotating 
 models with an improved multipole approximation of gravity
(e.g., \citet{couch13g}) explode more energetically than those only with 
 the monopole contribution (see, more details in Takiwaki et al. in preparation).}.
 Our fiducial 3D models are computed on a 
spherical polar grid with a resolution of $n_r \times n_{\theta} \times
 n_{\phi}$ = $384 \times 64 \times 128 $, in which non-equally
 spatial radial zones covers from the center to an outer boundary of
 5000 km.\footnote{This choice of the outer boundary position
 was shown to be insignificant especially 
in the simulation timescale ($\lesssim$ 300 ms postbounce)
  in this work (see section 2.3 
 in \citet{naka15}).}
 Our spatial grid has a finest mesh spacing $dr_{\rm min} = 0.5$ km
 at the center and $ dr/r$ is better than 2\% at $r \ge 100$ km.
 For a numerical resolution test, we compute high-resolution runs 
with $n_r \times n_{\theta} \times n_{\phi}$ = $384 \times 128 \times 256 $.

In total, we have computed nine 3D models, which 
 consists of six models with the fiducial resolution 
(i.e., the two progenitors with $\Omega_0 = 0, 1, 2$ rad/s) and 
  three high-resolution runs for the 11.2 $M_{\odot}$ model.
By using the fastest $K$ computer in Japan, it typically took 2 months 
(equivalently $\sim$ 15 Pflops-day computational resources) 
for each of the high-resolution runs.

\begin{figure}
\begin{center}
\includegraphics[width=0.99\linewidth]{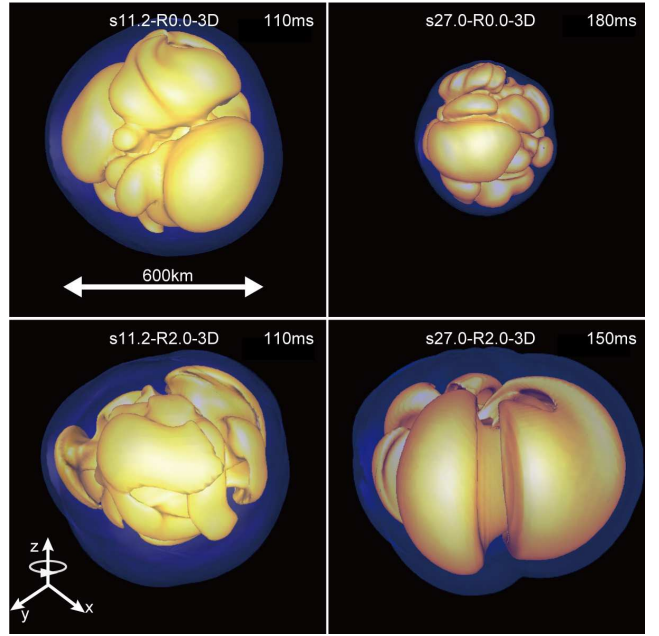}
\end{center}
\caption{3D iso-entropy surfaces showing
 the blast morphology for the non-rotating (top panels) and 
 rapidly rotating (bottom panels) models 
of the $11.2$ ({\it left}) and $27.0 M_{\odot}$ star ({\it right}), respectively.
 For each panel, the time is given at the top right corner, 
 which is measured relative to core bounce ($t \equiv 0$).
 The rotational axis is shown in the left bottom panel (z-axis) and the
 viewing angle of each plot is all the same.
}\label{f1}
\end{figure}

\section{Results}\label{sec:results}

Figure \ref{f1} summarizes the blast morphology for 
the $11.2 M_{\odot}$ (left panels) and $27.0 M_{\odot}$ star (right panels), 
which are helpful to compare the hydrodynamics features 
between the non-rotating ({\it top}) and rapidly rotating ({\it bottom}) 
models, respectively.

 In the non-rotating models, s11.2-R0.0-3D ({\it top left}) shows typical features of 
neutrino-driven convection in the postshock regions. The rising plumes
 grow stronger and larger in angular size from the initial small mushroom-like
Rayleigh-Taylor fingers. 
 
In models with rapid rotation,
 a clear oblate explosion is obtained for model s27.0-R2.0-3D
 ({\it bottom right}), in which the revived
 shock expands more strongly in the equatorial plane.
 This feature is only weakly visible for model s11.2-R2.0-3D ({\it bottom left})
 due to the early shock revival (see also, top panel of Figure 2).
 Later we present detailed analysis of the origin of the oblate 
explosion and point out a new aspect of rapid rotation for assisting 
explosions. 

\begin{figure}
\begin{center}
\includegraphics[width=.90\linewidth]{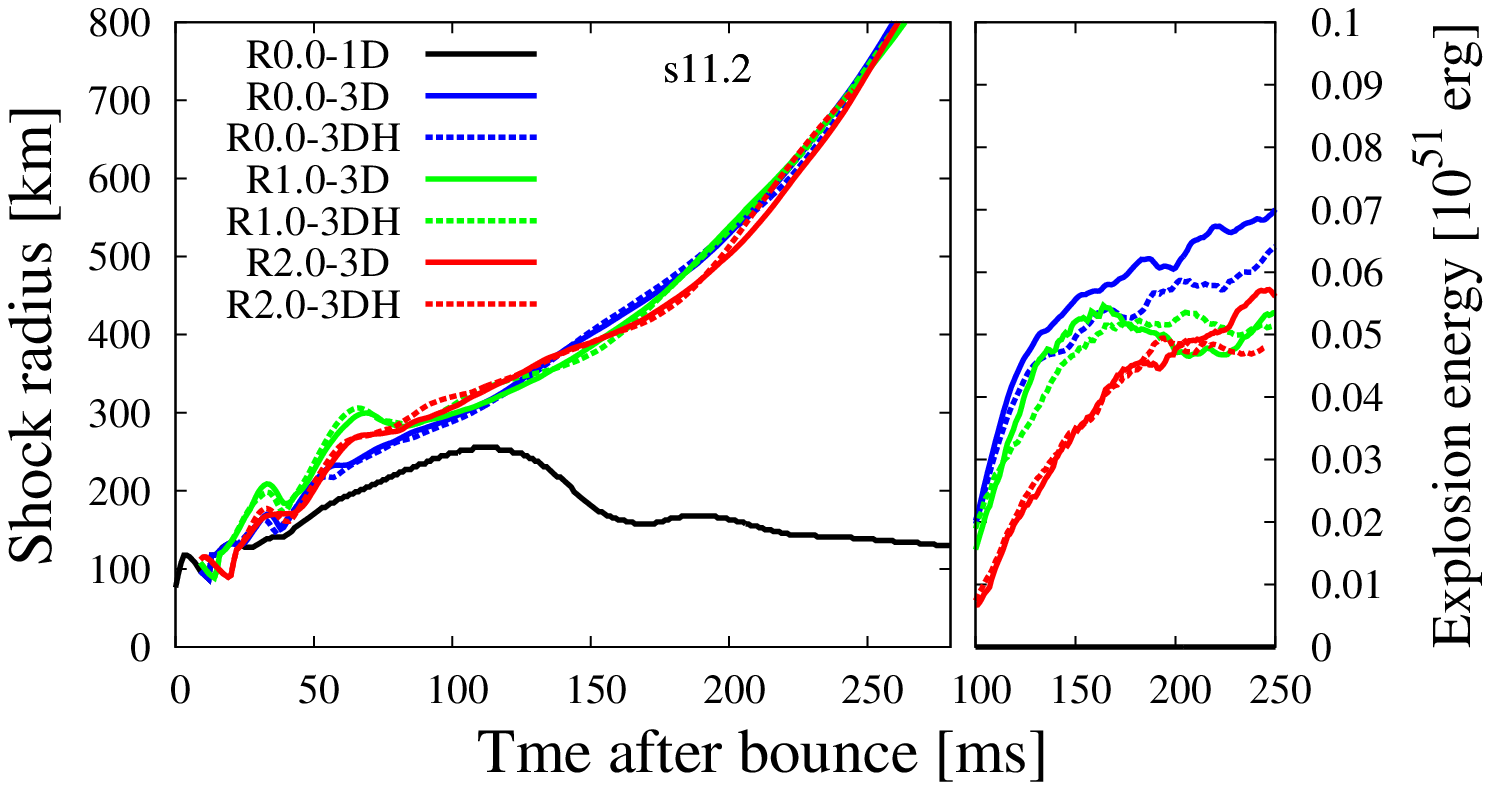}\\
\includegraphics[width=.90\linewidth]{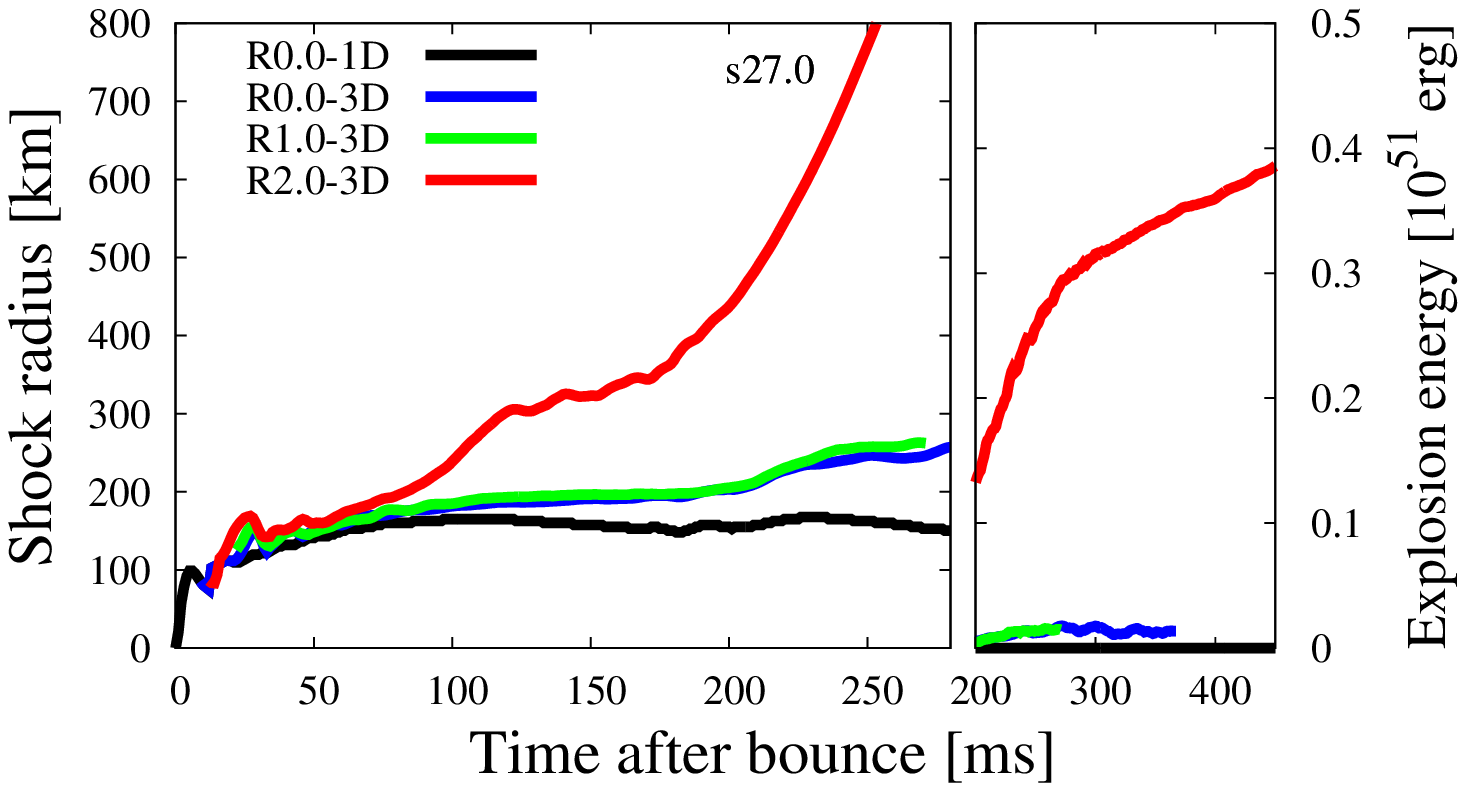}\\
\end{center}
\caption{Evolution of the (average) shock radii and the diagnostic explosion energy 
 for the 11.2 and 27 $M_{\odot}$ star, respectively.
In the top panel, the model name with ``H'', corresponds the high resolution model.}\label{fig:rshexp}
\end{figure}

 \begin{figure}
  \begin{center}
   \begin{minipage}{.99\linewidth}
    \includegraphics[width=.90\linewidth]{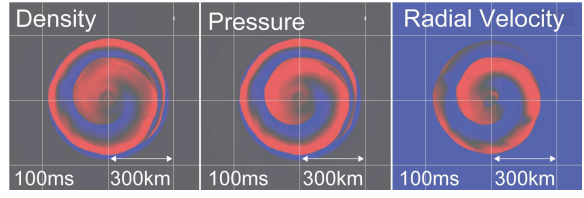}
    \subcaption{Deviations of the density (left panel), pressure (middle panel), and 
radial velocity (right panel) from the angle averaged value (see text) 
in the equatorial plane of s27.0-R2.0-3D at 100 ms postbounce.}\label{fig:3-a}
   \end{minipage}\\
  \begin{minipage}{.99\linewidth}
   \includegraphics[width=.90\linewidth]{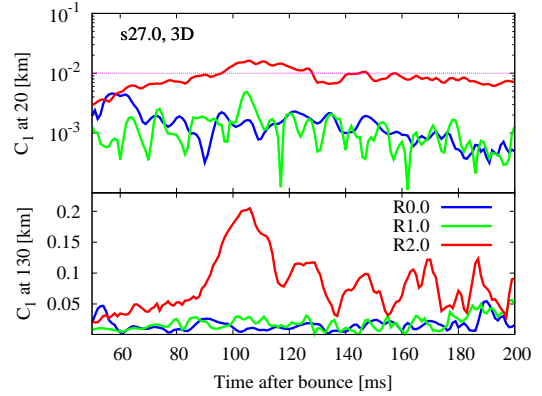}
    \subcaption{Evolution of dipole mass deformation (see
   text for the definition).  That is measured at 20 km and 130km for the top and the bottom panels, respectively.}\label{fig:3-b}
   \end{minipage}\\
  \begin{minipage}{.99\linewidth}
   \includegraphics[width=.90\linewidth]{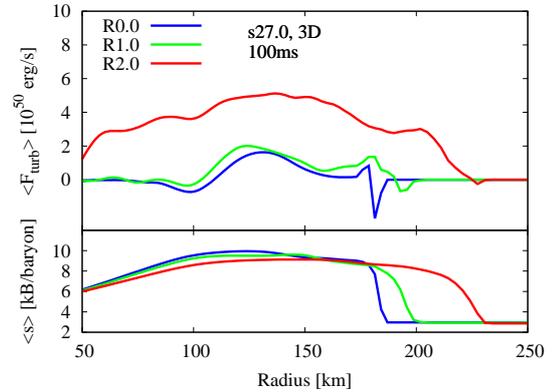}
    \subcaption{Radial profile of the turbulent energy flux (see Equation (1)) and 
entropy (per nucleon) at 100 ms postbounce.}\label{fig:3-c}
   \end{minipage}
\caption{Detailed analysis of the spiral flows and the resulting energy transport.}
 \end{center}
 \end{figure}

Before going into detail, 
let us shortly summarize the evolution of the shock and (diagnostic)
 explosion energy of all the computed models in Figure 2.
The top panels are for the 11.2 $M_{\odot}$ series with different $\Omega_0$
 and different numerical resolution (with the high resolution being ended with H).
 The average shock radii of the standard resolution models (solid
line) and high resolution models (dashed line) do not deviate from each
other. It is important to present that our results do not strongly depend on the 
grid size\footnote{Apparently our resolution is not sufficient for reproducing
realistic viscosity \citep{Couch15}). The convergence
 may be partly due to the diffusive feature of the HLLE scheme employed in this work
 (e.g., \citet{radice15}).}.

The bottom panel of Figure 2 shows that all the variations of the non-rotating 
27$M_{\odot}$ progenitor star do not trend toward an explosion very 
clearly during the simulation, 
whereas the rapidly rotating model does so (red solid line) 
with the diagnostic energy much bigger than those of the non-rotating models.
Using the same progenitor (27 $M_{\odot}$), the hydrodynamics features in
 the non-rotating models
 are qualitatively consistent with \citet{Hanke13}.

From here we proceed to focus on the impacts of rotation on the shock dynamics.
As shown in the top panel of Figure \ref{fig:rshexp},
 the shock radius of the rotating $11.2 M_{\odot}$ models (red and green) does not deviate from that of 
the non-rotating model (blue). Regarding the diagnostic energies, 
the rotational model shows less energetic explosion. 
That is because the neutrino luminosities of the rotating models are
slightly smaller than that of the non-rotating models (e.g., \citet{Marek09}).
In the case of such an early shock revival observed in the light progenitor,
the rotation does not help the onset of the explosion and 
even decrease the explosion energy.
It should be noted that time derivative of the diagnostic explosion energies, 
$E_{\rm diag}$,
are also small in these models and they seem to be almost saturated
 in spite of the relatively short postbounce time ($\sim 250$ ms).

For the heavier progenitor of 27.0$M_{\odot}$, the situation is inverted.
As shown in the bottom panel of Figure \ref{fig:rshexp},
only the rapidly rotating model explodes.
From here, we focus on the 27.0$M_{\odot}$ models since the effects of the rotation 
 is most distinct.
We try to provide a new interpretation of the rotating explosion.
For later convenience, let us define the deviation from the
spherical (angle-)averaged variable as $(A -\langle A\rangle)/\langle A\rangle$,
where $\langle A \rangle$ represents the angle average of $A$.
In the panels of Figure \ref{fig:3-a},
the deviation of the density, pressure, and radial velocity in the 
equatorial plane of R2.0-3D is shown, respectively.
The spiral flows (colored by red in the panels) are associated with 
high density and pressure, which pushes the matter outward and assists 
the explosion.

At $\sim$ 80 ms after bounce, the value of the rotation to the gravitational energy 
($T/|W|$) in the vicinity of the PNS exceeds 6\% and the iso-density surface of the PNS
 begins to be deformed with the dominance of $m = 1 $ mode. This behavior is 
quite similar to \citet{ott05} who were 
 the first to observe the growth of the low-$T/|W|$ instability in the PNS context.
Figure \ref{fig:3-b} shows the amplitude of 
the dipole mass deformation in the equatorial plane, whose definition is
$C_1(r)= \int_0^{2\pi} \rho Y_{1,1} \mathrm{d}\phi/\int_0^{2\pi} \rho Y_{0,0} \mathrm{d}\phi$, where $Y_{\ell,m}$ represents the spherical harmonics. After $\sim 90$ ms postbounce, the 
 deformation amplitude in the center ($C_1$ at 20 km, top panel of Figure \ref{fig:3-b}) approaches a percent level
 (as is consistent with \citet{ott05}),
 simultaneously, the spiral flows begin to extend outwards later on 
(e.g., left and center panels of Figure \ref{fig:3-a}).
The deformation amplitude near at the stalled shock 
($C_1$ at 130 km)
 peaks at around $100\sim 110$ ms postbounce with the maximum amplitude of 
 $\sim 20\%$ in density.
 It should be noted that the rotation energy of the mildly rotating model
 does not reach the threshold of the onset of low-$T/|W|$ instability.
 In this case
 the hydrodynamic evolution is similar to that in the non-rotating model.

The spiral flow is not merely ``pattern'' of the density and the pressure.
This wave really brings the mass and the thermal energy from the central 
to the outer region.
The right panel of Figure \ref{fig:3-a} shows the snapshot of the
radial velocity. In red region, matter has a positive radial velocity
and in the blue region vice versa.
From these three panels of Figure \ref{fig:3-a}, one can see that the matter with the 
high density and high thermal energy (pressure) goes outward and 
that with low density and low thermal energy goes inward.

To quantify the energy transport by the spiral flows, we utilize
 the concept of the (turbulent) energy flux 
as \begin{eqnarray}
\langle F_{\mathrm{turb}}\rangle &=& \langle (e_{\rm{int}}+p)^{\prime}v^{\prime}_r\rangle\label{eq:fturb},
\end{eqnarray}
 where $\rho, p, v_r$ represents the density, pressure, and the radial velocity
 respectively and the dashed values denote the deviation from the average.
 This flux means energy transport caused by instabilities in the 
context of Reynolds decomposition \citep{murphy11,murphy13}.
The kinetic energy can be ignored since that is  typically ten times smaller than the internal energy.
The total energy transport is written as the sum of one-dimensional
energy flux, $\langle e_{\rm{int}}+p \rangle \langle v_r \rangle$, and
the modification by the spiral flows in this case,  
$\langle (e_{\rm{int}}+p)^{\prime}v^{\prime}_r\rangle $,
since the background flows (without
 the spiral flows) yield to the following relations $\langle
(e_{\rm{int}}+p)^{\prime}\rangle =\langle v^{\prime}_r\rangle =0$.
Note that we actually estimate the flux as 
$\langle F_{\mathrm{turb}}\rangle \sim \langle \rho\left((e_{\rm{int}}+p)/\rho\right)^{\prime}v_r \rangle$ because this treatment empirically suppresses artificial overestimation 
 due to the steep density gradient near at the shock (e.g., \citet{murphy13}).

Figure \ref{fig:3-c} shows the energy flux and 
radial profile of the entropy. The blue and green line corresponds to the 
non-rotating model and mildly rotating model, respectively.
As was previously known in the non-rotating model,
the turbulent flux (blue line) makes a significant contribution to the energy 
transport only in the postshock region with negative entropy gradient (compare 
 two panels in Figure \ref{fig:3-c}). On the other hand, it is shown 
 for the rapidly rotating model (red line) that 
 the energy flux continuously contributes to the energy transport 
from the central region to behind the shock.

Although similar results are obtained by \citet{Ott12a}, 
the detailed mechanism was not discussed since the work was dedicated to the gravitational wave emission.
In addition, there are severe limitations in \citet{Ott12a}.
Since an octant symmetry was assumed in their work,
$m=1$ mode which is dominant in our model, was not resolved.
Their neutrino transport is based on the leakage scheme and 
the rotation aided explosion was obtained only in adiabatic models
whose neutrino cooling and heating is switched off (see their Figure 7).

\begin{figure}
\begin{center}
\includegraphics[width=.85\linewidth]{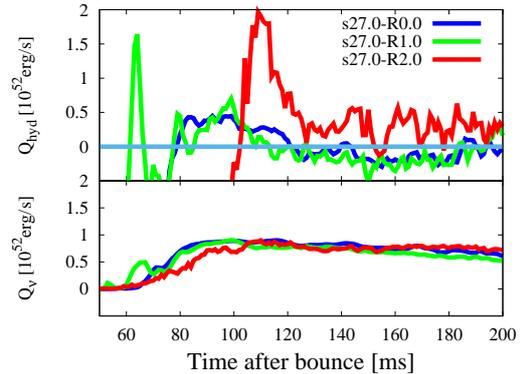}
\end{center}
\caption{Contribution of the energy gain (rate) due to the spiral activity 
 in the gain region (top panel, see Equation (2) for details) relative to the neutrino 
heating rate (bottom panel) 
for the non-rotating (blue line), mildly rotating 
(green line), and rapidly rotating (red line) 
27 $M_{\odot}$ models, respectively.
}\label{fig:qhyd}
\end{figure}

How much does that spiral activity power the shock ?
 To evaluate this, we estimate the rate of the energy change in the gain region
 ($r_{\rm gain} < r < r_{\rm shock}$) 
 as
\begin{eqnarray}
Q_{\mathrm{hyd}} &=& 
\int d \cos\theta d \phi
\left[ (\frac{1}{2}\rho v^2 + e + \rho \Phi)r^2v_r
\right]^{r=r_{\rm gain}}_{r=r_{\rm shock}}, \label{eq:qhyd}
\end{eqnarray}
 where $\Phi$ is the gravitational potential.
Note that $Q_{\mathrm{hyd}} $ can be clearly distinguished from the rate of the energy increased by the neutrino heating at that moment
(see eq. (107) of \citet{Janka01}  for more general description).
As shown in the top panel of Figure \ref{fig:qhyd}, $Q_\mathrm{hyd}$ for 
 the rapidly rotating model sharply grows 
 from zero at $\sim 100$ ms postbounce and peaks around $\sim 110$ ms postbounce. The 
 timescale coincides with the epoch when the developing spiral flow from the center 
 (Figure \ref{fig:3-b}) thrusts into the gain region.
 Note also that both of the time variability of $Q_{\rm hyd}$ (Figure \ref{fig:qhyd})
 and $C_{1}$ at the radius of 130 km (bottom panel of Figure \ref{fig:3-b}) is correlated.
 In the non-rotating or mildly rotating cases (blue and green lines), on the other hand,  
  $Q_\mathrm{hyd}$ takes negative value after $\sim$ 130 ms postbounce. In fact,
 the shock revival has not yet been obtained for these models in the simulation timescale.
These analyses naturally support that the rapidly rotating model trends toward an 
explosion because of the energy gain sustained by the spiral activity.

The energy gain due to the spiral activity is comparable to that of the neutrino heating.
The bottom panel of Figure \ref{fig:qhyd} shows the time evolution of 
$Q_\mathrm{\nu}$ that is an integrated neutrino heating rate in the gain region.
The value is $\sim 10^{52}$ erg/s in all the models and the contribution
of $Q_\mathrm{hyd}$ in the rapidly rotating model is the same order to that of the 
 net heating rate. The rapid rotating model is surely energized by the spiral flows.

\section{Summary and Discussion}\label{sec:summary}

We reported results from a series of 3D core-collapse simulations for 
$11.2$ and 27 $M_{\odot}$ stars using the IDSA scheme for spectral neutrino
 transport. By changing the initial strength of rotation systematically, 
we observed a new indication of rotation-assisted explosion 
for the $27.0 M_{\odot}$  model. This model fails to explode in the corresponding 
non- or mildly rotating models. In the rapidly
 rotating case, strong spiral flows generated
 by the low $T/|W|$ instability enhance
the energy transport from the PNS to the gain 
region, which makes the shock expansion more energetic.
These impacts of the rotation were also shown to be sensitive
 to the progenitor models. For the lighter progenitor, 
the early shock revival is little affected by the rotation.
In this case, the explosion energy of the rotating model becomes
 weaker, as previously pointed out by \citet{Marek09},
 predominantly due to the smaller neutrino luminosity.


The major limitation of this study would be omission of the magnetic fields. 
At present, saturation level of the field amplification due to magnetorotational 
instability (MRI) is still under a hot debate \citep{ramb16}.
 Depending on the MRI's growth rate,  
the magnetic fields could be amplified during our simulation time, possibly
 fostering the onset of explosion  \citep{moesta15,sawai13}. 
To test this, a high-resolution 3D MHD model is needed, which is another
 major undertaking.

The final (averaged) angular velocity of model s27-R2.0 is $\sim 2000\ \rm{rad/s}$
 in the vicinity of the PNS, which is apparently too fast to be reconciled with 
 observations of canonical radio pulsars (e.g., \citet{faucher}, see also 
\citet{ott06}). Such rapid rotation (e.g., $\Omega_0 = 2$ rad/s in this work), rare 
as it should be (see discussions in \citet{Woosley06,Cantiello07,fuller15,Chat}),
 is attracting much attention as to their possible relevance to hyper-energetic explosions (e.g., 
\citet{iwamoto98,mazzali08}). These events are also
hypothetically related to the formation of magnetars and collapsars (e.g.,
 \citet{meszaros06} for a review).
 We hope that this work could give a momentum for theorists
 to pay more attention to 3D models with rapid rotation (plus magnetic fields, e.g.,
 \citet{moesta15,masada15} for recent discoveries), 
which could possibly illuminate the yet unexplored variety of the explosion dynamics 
 where non-axisymmetric instabilities play a substantial role.
 

 \section*{Acknowledgements}
T.T. is grateful to K. Nitadori and J. Makino for the tuning of
our code. 
 The computations in this research were performed 
 on the K computer of
 the RIKEN AICS through 
the HPCI System Research project (Project ID:hp120304) together with 
XC30 of CfCA in NAOJ.
YS was supported by JSPS postdoctoral fellowships for research abroad.
This study was supported in part by the Grants-in-Aid for the 
Scientific Research from the Ministry of Education, Science and Culture of 
Japan (Nos. 24103006, 24244036,  26104001, 26707013, 26870823, 15H01039 and 15H00789) and by HPCI Strategic Program of Japanese MEXT.





\bibliographystyle{mnras}
\bibliography{mybib} 


\label{lastpage}
\end{document}